\documentclass[preprint,aps,pra,groupedaddress,superscriptaddress,floatfix]{revtex4-1}

\usepackage{graphicx}
\usepackage{nicefrac}
\usepackage{amsmath}
\usepackage{amsfonts}
\usepackage{amssymb}
\usepackage{amsthm}
\usepackage{epsf}
\usepackage{bm}
\usepackage{mathtools}
\usepackage{commath}
\usepackage{pgf}
\usepackage{fancyhdr}
\usepackage{subfigure}
\usepackage{epstopdf}
\usepackage{gnuplottex}

\usepackage{graphicx}
\graphicspath{{graph/}}
\DeclareGraphicsExtensions{.pdf,.png,.jpg, .eps}

\DeclarePairedDelimiter{\bra}{\langle}{\rvert}
\DeclarePairedDelimiter{\ket}{\lvert}{\rangle}

\allowdisplaybreaks[1]

\newcommand{\balpha}{\bm{\alpha}}

\newcommand{\bfF}{{\bf F}}
\newcommand{\bfI}{{\bf I}}
\newcommand{\bfJ}{{\bf J}}

\newcommand{\bfn}{{\bf n}}
\newcommand{\bfp}{{\bf p}}

\newcommand{\bfr}{{\bf r}}

\newcommand{\MJ}{{M_J}}
\newcommand{\Vnucl}{V_\mathrm{nuc}}
\newcommand{\Vmagn}{V_\mathrm{m}}
\newcommand{\Vhfs}{V_\mathrm{hfs}}
\newcommand{\muB}{\mu_\textrm{B}}

\newcommand{\Es}{E_{1s}}
\newcommand{\Egs}{E_{(1s)^2}}
\newcommand{\DEph}{\Delta E_\mathrm{1ph}}
\newcommand{\etal}{\textit{et al.}}
\newcommand{\matrixel}[3]{\langle #1 | #2 | #3 \rangle}

\newcommand{\Bmatrixel}[3]{\Big\langle #1 \Big| #2 \Big| #3 \Big\rangle}

\usepackage{color}
\definecolor{BLUE}{rgb}{0.0,0.0,1.0}
\newcommand{\rn}[1]{{#1}}
\usepackage{soul}
\soulregister\cite7
\soulregister\ref7
\newcommand{\rs}[1]{{}}

\begin{document}
% \setcounter{page}{2}
%  \setstretch{1.3}
%  \begin{titlepage}
%\pagestyle{plain}

\title{Helium-like ions in magnetic field: application of the nonperturbative relativistic method for axially symmetric systems}

\author{A.~M.~Volchkova}
\affiliation{Department of Physics, Saint-Petersburg State University, 199034 Saint-Petersburg, Russia}

\author{V.~A.~Agababaev}
\affiliation{Department of Physics, Saint-Petersburg State University, 199034 Saint-Petersburg, Russia}
\affiliation{Saint-Petersburg State Electrotechnical University ``LETI'', 197376 Saint-Petersburg, Russia}

\author{D.~A.~Glazov}
\affiliation{Department of Physics, Saint-Petersburg State University, 199034 Saint-Petersburg, Russia}

\author{A.~V.~Volotka}
\affiliation{Helmholtz-Institut Jena, D-07743 Jena, Germany}
\affiliation{GSI Helmholtzzentrum f\"ur Schwerionenforschung GmbH, D-64291 Darmstadt, Germany}

\author{S.~Fritzsche}
\affiliation{Helmholtz-Institut Jena, D-07743 Jena, Germany}
\affiliation{GSI Helmholtzzentrum f\"ur Schwerionenforschung GmbH, D-64291 Darmstadt, Germany}
\affiliation{Theoretisch-Physikalisches Institut, Friedrich-Schiller-Universit\"at Jena, D-07743 Jena, Germany}

\author{V.~M.~Shabaev}
\affiliation{Department of Physics, Saint-Petersburg State University, 199034 Saint-Petersburg, Russia}

\author{G.~Plunien}
\affiliation{Institut f\"ur Theoretische Physik, Technische Universit\"at Dresden, D-01062 Dresden, Germany}

% \begin{center}
% \textsc{\Large\textbf{Helium-like ions in magnetic field: application of the nonperturbative relativistic method for axially symmetric systems}}
% \end{center}
% \vspace{2.0em}
% \begin{center}
%     $A.~M.~Volchkova^{\ref{1}}\footnote{am.volchkova@gmail.com}$, $D.~A.~Glazov^{\ref{1}}$
    
%     \begin{enumerate}
%     \item
%     \label{1}
%     Department of Physics, St. Petersburg State University, Universitetskaya 7/9, 199034 St. Petersburg, Russia
%     \end{enumerate}
% \end{center}
% \begin{center}

\begin{abstract}
Dirac equation for an electron bound by a nucleus in the presence of external axially symmetric field can be solved numerically by using the dual-kinetic-balance conditions imposed on the finite basis set (A-DKB method [Rozenbaum \etal, Phys. Rev. A~\textbf{89}, 012514~(2014)]). We present the application of this method to describe helium-like ions exposed to homogeneous external magnetic field. The second-order Zeeman shift and the nuclear magnetic shielding constant are evaluated for the ground state, including the leading contribution of the interelectronic interaction. The A-DKB values are compared with the direct calculations by perturbation theory. The results for the nuclear magnetic shielding can serve for accurate determination of the nuclear magnetic moments. The quadratic contribution to the Zeeman effect can be relevant for high-precision measurements of the transition energies in helium-like ions.
\end{abstract}
\maketitle
% \textsc{\textbf{Abstract}}

% The method of dual kinetic balance for axially symmetric systems (A-DKB) is used for helium-like ions. The results for the quadratic contribution to the Zeeman effect and the nuclear magnetic shielding constant are presented. We also compared the results obtained using the A-DKB method with the results obtained by perturbation theory. The results obtained can be used to interpret experimental data. In particular, to obtain accurate values of the magnetic moments of nuclei.
% \end{center}
% 
% 
% \end{titlepage}
% 
\section {Introduction}
\label{sec:intro}

Bound-electron $g$ factor, which mostly determines Zeeman splitting in highly charged ions, is measured with increasing precision during the last two decades~\cite{sturm:17:a}. The relative experimental uncertainty has reached $2.4\times 10^{-11}$ in H-like carbon~\cite{sturm:14:n}, $0.7\times 10^{-10}$ in Li-like silicon~\cite{glazov:19:prl}, and $1.4\times10^{-9}$ in B-like argon~\cite{arapoglou:19:prl}. The $g$-factor measurements already performed and anticipated in the near future, combined with the corresponding theoretical efforts, provide access to the fundamental constants and nuclear properties~\cite{shabaev:15:jpcrd,harman:18:jpcs}. In particular, the nuclear magnetic moments can be determined with unprecedented precision from the $g$ factors of few-electron ions~\cite{werth:01:ha,quint:08:pra}. This task has become particularly relevant after a discrepancy was found between the recent measurement of the hyperfine splitting in H- and Li-like bismuth~\cite{ullmann:17:natcommun} and the most accurate theoretical prediction~\cite{volotka:12:prl}. This so-called ``hyperfine puzzle'' has been resolved with the new value of $^{209}$Bi nuclear magnetic moment~\cite{skripnikov:18:prl}, which disagrees with the previously accepted one. The more general outcome of this work is that the uncertainty of the magnetic moment values determined by the nuclear magnetic resonance method can be significantly underestimated. Recently, the new value for $^{207}$Pb has been determined~\cite{fella:20:prr} in strong disagreement with the tabulated value. The nuclear magnetic shielding for a bound electron in the $1s$ and some excited states was studied within the fully relativistic approach in Refs.~\cite{moore:99:mp,pyper:99:mp1,pyper:99:mp2}. Later, detailed theoretical investigations have been presented for the ground state of H-like~\cite{moskovkin:04:pra,moskovkin:06:pra,yerokhin:11:prl,yerokhin:12:pra}, Li-like~\cite{moskovkin:08:os,moskovkin:08:pra}, and B-like ions~\cite{volchkova:17:nimb}. In this work, we study the nuclear magnetic shielding for the ground state of He-like ions. The total magnetic moment is fully determined by the nucleus and the shielding constant in this case. Despite certain experimental difficulties, in principle, this allows one to access directly the nuclear magnetic moment in high-precision Penning-trap measurements.

The nonlinear contributions to the Zeeman splitting can play an important role in high-precision measurements. In particular, the second- and third-order effects can be detected in the Penning-trap experiments with B-like ions~\cite{lindenfels:13:pra}. Recent measurement of the ground-state $g$ factor in $^{40}$Ar$^{13+}$~\cite{arapoglou:19:prl} was sensitive to the third-order contribution~\cite{glazov:13:ps,varentsova:17:nimb,varentsova:18:pra}. Subtraction of the second-order contribution~\cite{glazov:13:ps,agababaev:17:nimb,varentsova:18:pra,agababaev:20} was required to obtain the most precise up-to-date experimental value of the fine-structure transition energy in B-like argon~\cite{egl:19:prl}. The nonlinear Zeeman effects are enhanced by the closely spaced levels of the same parity --- $^2P_{1/2}$ and $^2P_{3/2}$ in B-like ions --- which are mixed by the external magnetic field. Similar situation can take place for $n=2$ levels in low- and middle-$Z$ He-like ions. So, the quadratic Zeeman shift can be relevant for future high-precision measurements. Transition energies in He-like ions serve as a perfect probe of the many-electron QED effects and therefore attract much experimental and theoretical interest, see, e.g., Refs.~\cite{epp:15:pra,beiersdorfer:15:pra,machado:18:pra,malyshev:19:pra,kozhedub:19:pra,yerokhin:19:jpcrd} and references therein. In this work, we investigate the second-order Zeeman effect for the ground state of low- and middle-$Z$ He-like ions. Consideration of the excited states of He-like ions will be the subject of our future work.

Both the nuclear magnetic shielding and the quadratic contribution in magnetic field represent the terms of the second order of perturbation theory. Here we demonstrate the application of the nonperturbative method to solve the Dirac equation in the presence of axially symmetric field~\cite{rozenbaum:14:pra}. The sought-for contributions are found by taking the derivative of the energy with respect to the field strength. In addition to the one-electron part, we consider the first-order interelectronic-interaction correction. The results obtained within this method are tested against the straightforward perturbation-theory calculations. This method has significant advantages for evaluation of the higher-order contributions, which can be too cumbersome within the perturbation theory.

Relativistic units ($\hbar = 1$, $c = 1$, $m_e = 1$) and Heaviside charge unit [$\alpha=e^2/(4\pi)$, $e<0$] are employed throughout the paper, $\muB=|e|/(2m_e)$ denotes the Bohr magneton, $m_p$ is the proton mass, and $m_e$ is written for clarity in the electron-to-proton mass ratio $m_e/m_p$.

\section {\rs{A-DKB method}\rn{Dirac equation for axially symmetric potential}}
\label{sec:A-DKB}

Consider the stationary Dirac equation\rs{:}\rn{,}
\begin{equation}
\label{eq:D} 
  \hat H \psi(\bfr) = E \psi(\bfr) 
\;,
\end{equation}
\rn{with the Hamiltonian,}
\begin{equation}
\label{eq:hD} 
  \hat{H} = \balpha \cdot \bfp + \beta + \hat{V}
\;,
\end{equation}
\rn{where $\balpha$ and $\beta$ are the Dirac matrices, $\hat{V}(r,\theta)$ is an arbitrary axially symmetric binding potential. Throughout the paper, we consider the electron bound by the nucleus in the presence of magnetic field. The symmetry axis of this field determines the $z$-axis of the spherical coordinate system $(r,\theta,\phi)$. So, $\hat{V}(r,\theta)$ is a multiplication operator with some matrix structure, it will be given explicitly in Section~\ref{sec:Zeeman}.}
\rs{here} 
% \[
% V = V_{\mathrm {nucl}}(r) + V_{\mathrm {ext}}(r,\theta)\;,
% \]
\rs{where
$V_{\mathrm {nucl}}(r)$ --- nuclear potential,
$V_{\mathrm {ext}}(r,\theta)$ --- axial symmetrical external potential.}

To solve this equation \rn{numerically}, we use the A-DKB method \rn{developed in Ref.~\cite{rozenbaum:14:pra}.} \rs{This method is} \rn{It represents} a generalization of the dual-kinetic-balance (DKB) method \cite{shabaev:04:prl} for axially symmetric systems. \rs{Method A-DKB is based on described in detail in the article \cite{rozenbaum:14:pra}, here we give the main features.}
\rn{These methods are based on the finite-basis-set decomposition of the wave function, while the DKB conditions prevent emergence of the spurious states. It was demonstrated in Ref.~\cite{rozenbaum:14:pra} that this method reproduces the one-electron binding energies for an atom in the presence of external homogeneous electric or magnetic field in a wide range of the field strengths. Below we briefly describe the key elements of the A-DKB method.}

\rs{Such a Hamiltonian, unlike the Hamiltonian of a spherically symmetric field} \rn{In contrast to the spherically symmetric case, the Hamiltonian (\ref{eq:hD})} does not commute with the square of \rs{the operator of} the total \rs{orbital} \rn{angular} momentum.\rs{, but at the same time} \rn{Still}, commutation with the $z$-component \rs{of this operator} is preserved,
\[
[\hat {H}, \hat{J^2}] \neq 0, \quad [\hat H,\hat J_z]=0
\;.
\]
This means that \rn{only the $\varphi$-dependence can be separated explicitly. So,} the wave functions obtained from the stationary Dirac equation \rn{(\ref{eq:D})} \rs{with such a Hamiltonian} will be expressed as follows\rn{,}\rs{. Only the dependence on the variables $ \varphi $ can be separated.}
\begin{equation}
\label{eq:psi}
\psi(\vec r)=\frac{1}{r}
\begin{pmatrix}
G_1(r,\theta)e^{i\varphi(\MJ-\frac{1}{2})}\\
G_2(r,\theta)e^{i\varphi(\MJ+\frac{1}{2})}\\
iF_1(r,\theta)e^{i\varphi(\MJ-\frac{1}{2})}\\
iF_2(r,\theta)e^{i\varphi(\MJ+\frac{1}{2})}
\end{pmatrix}\;,
\end{equation}
\rn{where $\MJ$ is the eigenvalue of $J_z$. Equation (\ref{eq:D}) is then reduced to $H_\MJ \Phi = E \Phi$}
% 
% \begin{equation}
% \label{eq:Dr}
%   H_\MJ \Phi = E \Phi
% \end{equation}
% 
\rn{with} 
\begin{align}
\label{eq:Dr}
  H_\MJ &= 
  \begin{pmatrix}
     1     &  D_\MJ \\
    -D_\MJ & -1 \\
  \end{pmatrix}
    + \hat{V}
\\
  D_\MJ &= \left(\sigma_z \cos{\theta} + \sigma_x \sin{\theta}\right)\left( \frac{\partial}{\partial r} - \frac{1}{r}\right) 
\nonumber\\
    &+ \frac{1}{r}\left(\sigma_x \cos{\theta} - \sigma_z \sin{\theta}\right)\frac{\partial}{\partial \theta} + \frac{1}{r \sin{\theta}}\left( i\MJ \sigma_y + \frac{1}{2} \sigma_x \right)
\;.
\end{align}
\[
\]
\rs{To find the following functions}
\rn{The 4-component wave functions}
\begin{equation}
\label{eq:F}
  \Phi(r, \theta) = 
  \begin{pmatrix}
    G_1(r,\theta)\\
    G_2(r,\theta)\\
    F_1(r,\theta)\\
    F_2(r,\theta)
  \end{pmatrix}
\end{equation}
\rs{and their corresponding energies, the finite basis set method is used. Within this method, the functions (\ref{eq:F})} are decomposed into a finite basis set:
\begin{equation}
\label{eq:Phi}
\Phi(r, \theta) \cong \sum_{u=1}^4 \sum _{i_r=1}^{N_r} \sum_{i_{\theta}=1}^{N_{\theta}} C_{i_r {i_\theta}}^u W_{i_r i_\theta}^{u} (r, \theta) 
\end{equation}
\begin{equation}
 W_{i_r i_{\theta}}^{u} (r, \theta) = \Lambda B_{i_r} (r) Q_{i_{\theta}} (\theta) e_u \;,
\end{equation}
where
\[
e_1=
\begin{pmatrix}
1\\0\\0\\0
\end{pmatrix} , \qquad
e_2=
\begin{pmatrix}
0\\1\\0\\0
\end{pmatrix} , \qquad
e_3=
\begin{pmatrix}
0\\0\\1\\0
\end{pmatrix} , \qquad
e_4=
\begin{pmatrix}
0\\0\\0\\1
\end{pmatrix} \;.
\]
\(\{B_{i_r}(r)\}_{i_r=1}^{N_r}\) \rs{---}\rn{are the} \(B\)-splines of some order \(k\) \rn{and} \(Q_{i_{\theta}}(\theta)\) \rs{---}\rn{are the} Legendre polynomials: 
\[
 Q_{i_{\theta}}(\theta) = P_{i_{\theta}-1}\left(\frac{2}{\pi}\theta - 1\right)\;.
\]
\rn{The matrix $\Lambda$ enforces the DKB conditions~\cite{shabaev:04:prl,rozenbaum:14:pra},}
\[
\Lambda = 
\begin{pmatrix}
  1 & -\frac{1}{2}D_\MJ \\
  -\frac{1}{2}D_\MJ & 1
\end{pmatrix}
\;.
\]
To find the coefficients \(C_{i_r {i_\theta}}^u\) and the \rn{corresponding} energies, the expansion (\ref{eq:Phi}) is substituted into the Dirac equation and the generalized eigenvalue problem is solved numerically. The few lowest positive-energy states reproduce the bound states of Eq.~(\ref{eq:D}). The total set of the solutions forms the quasi-complete finite basis set, which can be used, in particular, to construct the Green function of Eq.~(\ref{eq:D}).

\subsection {\rn{Matrix elements of the magnetic and hyperfine interactions}}
\label{sec:me1}

Given the solutions of the Dirac equation, one can evaluate the matrix elements of various operators,
\begin{equation}
  \bra a \hat {F} \ket b =\int\limits_0^{\infty} dr\,r^2\, \int\limits_0^\pi d\theta\,\sin{\theta}\,\int\limits_0^{2\pi}d\varphi\,{\psi_a}^{\dagger}(r, \theta,\varphi) \,\hat{F}\, \psi_b (r, \theta,\varphi)
\;.
\end{equation}
The form of the wave functions~(\ref{eq:psi}) suggests that the integral over $\varphi$ is found analytically, and the integrals over $r$ and $\theta$ are calculated only numerically. Below we present the corresponding formulae.

The simplest case is the multiplication operator with the trivial matrix structure,
\begin{equation}
\label{eq:a-f-b}
  \bra{a}f(r, \theta)\ket{b}= \delta_{M_a, M_b}\int\limits_0^{\infty}\, dr\int\limits_0^\pi d\theta\,\sin{\theta}\, f\,(r, \theta) \, \left(G_1^aG_1^b+G_2^aG_2^b+F_1^aF_1^b+F_2^aF_2^b\right)
\;.
\end{equation}
Here and in the following, the upper index of the wave-function components indicate the state, $a$ or $b$, the wave-function arguments, $r$ and $\theta$, are implied. The $\delta$-symbol reflects the conservation of the angular momentum projection in this case.

In this work, we consider the interactions with the homogeneous external magnetic field and with the nuclear magnetic moment. The corresponding operators can be represented as $r^k\,[\bfn\times\balpha]_z$, where $\bfn=\bfr/r$. The matrix elements are expressed as follows,
\begin{multline}
\label{eq:a-fntaz-b}
  \bra{a}f(r, \theta)\, [\bfn\times\balpha]_z\ket{b}=\\= \delta_{M_a,M_b}\,\int\limits_0^{\infty}\, dr\int\limits_0^\pi d\theta\, \sin{\theta}\, f(r, \theta) \sin\theta\, \left( G_1^aF_2^b-G_2^aF_1^b-F_1^aG_2^b+F_2^aG_1^b\right)
\;.
\end{multline}
The two remaining components (in spherical notations: $a_{+}=a_x+ia_y$, $a_{-}=a_x-ia_y$) are
\begin{multline}
\label{eq:a-fnta+-b}
  \bra{a}f(r, \theta)\,[\bfn\times\balpha]_+ \ket{b}=
\\
  = \delta_{M_a,M_b+1}\,\int\limits_0^{\infty}\, dr\int\limits_0^\pi d\theta\, \sin{\theta}\, f(r, \theta) \left[\sin\theta\, \left(G_1^aF_1^b-G_2^aF_2^b-F_1^aG_1^b+F_2^aG_2^b\right)\right.
\\
  + \left.2\cos\theta \,\left(-G_1^aF_2^b+F_1^aG_2^b\right)\right]
\;,
\end{multline}
\begin{multline}
\label{eq:a-fnta--b}
  \bra{a}f(r, \theta)\, [\bfn\times\balpha]_-\ket{b}=
\\
  = \delta_{M_a,M_b-1}\,\int\limits_0^{\infty}\, dr\int\limits_0^\pi d\theta\, \sin{\theta}\, f(r, \theta) \left[\sin\theta\, \left(-G_1^aF_1^b+G_2^aF_2^b+F_1^aG_1^b-F_2^aG_2^b\right)\right.
\\
  + \left.2\cos\theta\, \left(G_2^aF_1^b-F_2^aG_1^b\right)\right]
\;.
\end{multline}
We also present for completeness the following formula for the scalar product $(\bfn\cdot\balpha)$,
\begin{multline}
\label{eq:a-fnda-b}
  \bra{a} f(r, \theta)\, (\bfn\cdot\balpha)\ket{b}=
\\
  = \delta_{M_a,M_b}\int\limits_0^{\infty}\, dr\int\limits_0^\pi d\theta\, \sin{\theta}\, i\,f(r, \theta)\, \left[\cos\theta \,\left(G_1^aF_1^b-G_2^aF_2^b-F_1^aG_1^b+F_2^aG_2^b\right)\right.
\\
  + \left.\sin\theta \, \left(G_1^aF_2^b+G_2^aF_1^b-F_1^aG_2^b-F_2^aG_1^b\right)\right]
\;.
\end{multline}
The formulae for the matrix elements containing the matrix $\beta$ can be obtained from the ones without $\beta$ (Eqs.~(\ref{eq:a-f-b})--(\ref{eq:a-fnda-b})) by changing the sign of the $F_{1,2}$-components of one of the wave functions, depending on the position of $\beta$.

% \begin{multline}
% \bra{a} f(r, \theta)\beta\,[\bfn\times\balpha]_z\ket{b}=\\= \delta_{M_a,M_b}\int\limits_0^{\infty}\, dr\int\limits_0^\pi \sin{\theta}\, d\theta\, r\,f(r, \theta)\, \sin\theta \,( G_1^aF_2^b-G_2^aF_1^b+F_1^aG_2^b-F_2^aG_1^b)\;.
% \end{multline}

% \begin{multline}
% \bra{a} f(r, \theta)\beta\,[\bfn\times\balpha]_+ \ket{b} = \\=\delta_{M_a,M_b+1}\int\limits_0^{\infty}\, dr\int\limits_0^{\pi} \sin{\theta}\, d\theta\, r\, f(r, \theta)\,(\sin\theta\, (G_1^aF_1^b-G_2^aF_2^b+F_1^aG_1^b-F_2^aG_2^b)+\\+\cos\theta \,(2G_1^aF_2^b+2F_1^aG_2^b))\;,
% \end{multline}

% \begin{multline}
% \bra{a} f(r, \theta)\beta\,[\bfn\times\balpha]_-\ket{b} =\\=\delta_{M_a,M_b-1}\int\limits_0^{\infty}\, dr\int\limits_0^\pi i \sin{\theta}\, d\theta \, r \,f(r, \theta)\,(\sin\theta\, (G_1^aF_1^b-G_2^aF_2^b-F_1^aG_1^b+F_2^aG_2^b)+\\+\cos\theta\, (2G_2^aF_1^b+2F_2^aG_1^b))\;,
% \end{multline}

% \begin{multline}
% \bra{a} f(r, \theta)\, \beta\, (\bfn\cdot\balpha)\ket{b}=\\=\delta_{M_a,M_b}\int\limits_0^{\infty}\, dr\int\limits_0^\pi \sin{\theta}\, d\theta \, r\, i\,f(r, \theta)\, (\cos\theta \,(G_1^aF_1^b-G_2^aF_2^b+F_1^aG_1^b-F_2^aG_2^b)+\\+\sin\theta\, (G_1^aF_2^b+G_2^aF_1^b+F_1^aG_2^b+F_2^aG_1^b))\;,
% \end{multline}

% Here components $+$ and $-$ are complex combinations:\\
% \[
% F_{+} = F_x+iF_y
% , \qquad
% F_{-} = F_x-iF_y.
% \]

Finally, we consider one particular nonlocal operator which is quite important for applications.
In the homogeneous external magnetic field, the parity of the state is a conserved quantum number, in addition to $\MJ$,
% Since the Hamiltonian commutes with the parity operator
\begin{align}
  [\hat H , \hat P]=0
\;,
\qquad
% \\
  \hat P \psi = \pm \psi
\;.
\end{align}
The spherical coordinates are transformed by the operator $\hat{P}$ in the following way,
\begin{equation}
  r \to r 
\;,\quad 
  \varphi \to \pi+\varphi 
\;,\quad 
  \theta \to \pi - \theta
\;.
\end{equation}
Additionally, $\hat{P}$ acts on the Dirac wave function with the matrix $\beta$~\cite{akhiezer}. So, the average value of $\hat{P}$ is given by,
\begin{multline}
\label{eq:a-P-a}
  \bra{a} \hat P \ket{a}=
   \int\limits_0^{\infty} dr\, r^2\, \int\limits_{0}^{\pi} d\theta\,\sin{\theta}\,\int\limits_0^{2\pi} d\varphi\,{\psi_a}^{\dagger}(r, \theta, \varphi) \beta \psi_a (r, \pi - \theta, \pi + \varphi)
\\
   =\int\limits_0^{\infty} dr\,\int\limits_{0}^{\pi} d\theta\,\sin{\theta}\,\int\limits_0^{2\pi}d\varphi \left( G_1(r,\theta)G_1(r, \pi-\theta) e^{i(\MJ-\frac{1}{2})\pi}+G_2(r,\theta)G_2(r, \pi-\theta) e^{i(\MJ+\frac{1}{2})\pi}\right.\\\left.-F_1(r,\theta)F_1(r, \pi-\theta) e^{i(\MJ-\frac{1}{2})\pi}-F_2(r,\theta)F_2(r, \pi-\theta) e^{i(\MJ+\frac{1}{2})\pi}\right)\\
   =e^{i\left(\MJ-\frac{1}{2}\right)\pi}\int\limits_0^{\infty}\, dr\int\limits_0^\pi d\theta\, \sin{\theta}\left(G_1(r,\theta)G_1(r, \pi-\theta) + G_2(r,\theta)G_2(r, \pi-\theta) e^{i\pi}\right.\\\left.-F_1(r,\theta)F_1(r, \pi-\theta) -F_2(r,\theta)F_2(r, \pi-\theta) e^{i\pi}\right)
\\
    =e^{i\left(\MJ-\frac{1}{2}\right)\pi}\int\limits_0^{\infty}\, dr\int\limits_0^\pi d\theta\, \sin{\theta}\, 
    \left( G_1\tilde{G}_1-G_2\tilde{G}_2
% \right.
% \\
%   \left.
    -F_1\tilde{F}_1+F_2\tilde{F}_2 \right)
\;,
\end{multline}
where $\tilde{G}_{1,2} = G_{1,2}(r,\pi-\theta)$, $\tilde{F}_{1,2} = F_{1,2}(r,\pi-\theta)$.
Evaluation of $\matrixel{a}{\hat{P}}{a}$, which may be equal $\pm 1$ only, allows us to identify the ``odd'' and ``even'' states in the spectrum.
% Thus, we can use the parity of the function to uniquely determine the energy level. To determine the parity of the wave function \(a\), it is necessary to consider the matrix element \(\bra a \hat P \ket a\). It can take two values --- $\pm 1$.
% then the solutions of the Dirac equation have a certain parity
% \begin{equation}
% \end{equation}
% although they do not have quantum numbers \(j\) and \(l\).\\

\subsection {\rn{Matrix elements of the} interelectronic interaction}
\label{sec:me2}

\rs{Interelectron interaction can be taken into account using one-photon exchange. In order to obtain a correction to the energy for one-photon exchange, it is necessary to sum the contributions of all possible diagrams. The correction to energy can be written in the form of the expression:}
\rn{The interaction between electrons can be considered within the perturbation theory (PT) or within any all-order method. In this work, we consider the ground state of helium-like ion and restrict ourselves to the first order of PT,
\begin{align}
\label{eq:Egs}
  \Egs = 2\,\Es + \DEph + \dots
\;.
\end{align}
% 
% where the second- and higher-order terms of the electron-electron interaction are not considered in this work.
% 
% . For the ground state of helium-like ion, the total energy can be written as
% For the two-electron system under consideration this contribution is written as
The one-photon-exchange correction is written as}
\begin{equation}
\label{eq:E_1ph}
  \DEph = \matrixel{a b}{I(0)}{a b} - \matrixel{b a}{I(\Delta_{ab})}{a b} 
\;.
\end{equation}
\rn{Here, $a$ and $b$ are the electron states, $\Delta_{ab}=E_a-E_b$ is zero in the present case, $I$ is the interaction operator, in the Feynman gauge it is given by}
\rs{Here $I$ is the one-photon exchange operator, $a$ is the valence electron, $b$ is the electrons of the closed shell, and summation is performed over them.
One-photon exchange operator in Feynman gauge:}
\begin{equation}
  I (\omega,r_{12}) = \alpha (1-{\balpha_1} \cdot \mathbf{\balpha_2}) \,\frac{\exp(i |\omega| r_{12})}{r_{12}}
\;.
\end{equation}
%where \(\omega = E_1-E_3\) (Fig.~\ref{ris:1ph}).
%\begin{figure}[h]
%\center{\includegraphics[width=0.2\linewidth]{1ph} }
%\caption{One-photon exchange diagram. A double line denotes a bound electron, a wavy line is a photon.}
% \label{ris:1ph}
%\end{figure}
\rn{This function of $r_{12}=|\mathbf{r}_1-\mathbf{r}_2|$ is represented by the following expansion in the spherical harmonics of $\mathbf{n_1}=\mathbf{r}_1/r_1$ and $\mathbf{n_2}=\mathbf{r}_2/r_2$,}
\rs{To calculate the correction to energy, it is necessary to calculate the matrix elements on the wave functions obtained from the Dirac equation in an axially symmetric field.}
\begin{equation}
\label{eq:exp-Y_lm}
  \frac{\exp(i \omega r_{12})}{r_{12}} = 4\pi \sum_{lm}^{\infty}\frac{1}{2l+1}g_l(\omega, r_1, r_2) Y^*_{lm}(\mathbf{n_2}) Y_{lm}(\mathbf{n_1})
\;,
\end{equation}
where
\begin{align*}
  g_l (\omega\neq 0,r_1,r_2) &= i \omega (2l+1)j_l(\omega r_{<})h_l^{(1)}(\omega r_>)
\;,\\
  g_l (0,r_1,r_2) &= \frac{r_<^l}{r_>^{l+1}}
\;,\qquad
  r_<=\min(r_1,r_2)
\;,\qquad
  r_>=\max(r_1,r_2)
\;.
\end{align*}
\rn{$j_l(x)$ and $h_l^{(1)}(x)$ are the spherical Bessel and Hankel functions, respectively. In Eq.~(\ref{eq:exp-Y_lm}) and in the following, the summation runs in the ranges $l=0,...\infty$ and $m=-l,...,l$. Use of the expansion (\ref{eq:exp-Y_lm}) allows for analytical integration over $\theta_{1,2}$ and $\phi_{1,2}$ in the spherically symmetric case. In the axially symmetric case only the $\phi$-integration can be performed analytically. The resulting expressions for the matrix elements are:} 
\rs{The integral over the angle $\varphi$ can be taken analytically:}
% 
% \begin{multline}
% \bra{a b} I_{\mathrm {F}} \ket { c  d} = \alpha \bra{ a  b} \frac{1-\mathbf{\alpha_1} \cdot \mathbf{\alpha_2}}{r_{12}} \mathrm {exp}(i |\omega| r_{12})\ket { c  d}  = \\\alpha \left(\bra{ a b} \frac{1}{r_{12}}\mathrm {exp}(i |\omega| r_{12}) \ket { cd}-\bra{ a b} \frac{\mathbf{\alpha_1} \cdot \mathbf{\alpha_2}}{r_{12}} \mathrm {exp}(i |\omega| r_{12})\ket {c d}\right) 
% \end{multline}
% We consider two terms individually:
% 
\begin{multline}
\label{eq:ab-IC-cd}
%   \bra{ a b} \frac{\exp(i \omega r_{12})}{r_{12}} \ket { c d} 
  \Bmatrixel{a b}{\frac{\exp(i \omega r_{12})}{r_{12}}}{c d} 
  = 
\\
    = \sum_{lm}^{\infty} \delta_{m,m_a-m_c} \delta_{m, m_d-m_b} \int\limits_0^{\infty} dr_1 \int\limits_{0}^{\pi} d\theta_1\, \sin\theta_1\, \Theta_{lm}(\theta_1) \int\limits_0^{\infty} dr_2 \int\limits_{0}^{\pi} d\theta_2\, \sin\theta_2\, \Theta_{lm}(\theta_2)\, g_l (\omega,r_1,r_2) 
\\ 
  \times \left( G_1^a G_1^c + G_2^a G_2^c + F_1^a F_1^c + F_2^a F_2^c \vphantom{G_1^b}\right) \left( G_1^b G_1^d + G_2^b G_2^d + F_1^b F_1^d + F_2^b F_2^d \right)
\;,
\end{multline}
\begin{multline}
\label{eq:ab-IB-cd}
  \Bmatrixel{a b}{\balpha_1 \cdot \balpha_2 \frac{\exp(i \omega r_{12})}{r_{12}}}{c d} = 
\\ 
    = \sum_{lm}^{\infty} \int\limits_0^{\infty} dr_1 \int\limits_{0}^{\pi} d\theta_1 \, \sin\theta_1\, \Theta_{lm}(\theta_1) \int\limits_0^{\infty} dr_2 \int\limits_{0}^{\pi} d\theta_2 \, \sin\theta_2\, \Theta_{lm}(\theta_2)\, g_l (\omega,r_1,r_2)
\\ 
    \times \left[2\delta_{m,m_a-m_c-1} \delta_{m, m_d-m_b-1} \left( G_1^aF_2^c - F_1^aG_2^c \vphantom{G_1^b}\right) \left( G_1^bF_2^d - F_1^bG_2^d \right)\right.
\\
    +2 \delta_{m,m_a-m_c+1} \delta_{m, m_d-m_b+1} \left( G_2^aF_1^c - F_2^a G_1^c \vphantom{G_1^b}\right)\left( G_2^bF_1^d - F_2^b G_1^d \right) 
\\
    + \left.\delta_{m,m_a-m_c} \delta_{m, m_d-m_b} \left( G_1^a F_1^c + F_2^a G_2^c - G_2^a F_2^c - F_1^a G_1^c \vphantom{G_1^b}\right)\left( G_1^b F_1^d + F_2^b G_2^d - G_2^b F_2^d - F_1^b G_1^d \right)\right]
\;,
\end{multline}
where
\[
  \Theta_{lm}(\theta) = \sqrt{\frac{(l-m)!}{(l+m)!}} P_l^m(\cos\theta)
\;.
\]
The wave-function arguments omitted in Eqs.~(\ref{eq:ab-IC-cd}) and (\ref{eq:ab-IB-cd}) for brevity reasons are $(r_1,\theta_1)$ for $a$ and $c$ states and $(r_2,\theta_2)$ for $b$ and $d$ states.

% In these calculations we use the following series expansion:
% \begin{equation}
% \frac{1}{r_{12}} \mathrm{exp} (i |\omega| r_{12}) = 4\pi \sum_{lm}^{\infty}\frac{1}{2l+1}g_l(\omega, r_1, r_2) Y^*_{lm}(\mathbf{n_2}) Y_{lm}(\mathbf{n_1})
% \end{equation}
% Here
% \[
% g_l (\omega,r_1,r_2) = i \omega (2l+1)j_l(\omega r_{<})h_l^{(1)}(\omega r_>),\
% \]
% \[
% g_l (0, r_1, r_2) = \frac{r_<^l}{r_>^{l+1}}.\
% \] 
% $r_<$ --- the smallest of $r_1$,$r_2$,\quad $r_>$ --- the greatest of $r_1$,$r_2$,\\
% $Y_{lm}$ --- spherical harmonics.

\rs{It should be noted that summation over l is infinite (in the case of a spherically symmetric field summation is finite).}
\rn{In the spherically symmetric field the summation over $l$ is truncated due to the angular algebra. In the present case, the summation remains infinite, although it is rapidly convergent for the systems and effects under consideration.} This fact, combined with the \rn{numerical integration over $\theta$}\rs{addition of the theta integral}, makes \rs{calculating}\rn{the} matrix elements \rn{calculation} much more time-consuming.

\section {Higher-order contributions to the Zeeman splitting}
\label{sec:Zeeman}
In this section, we consider various aspects of the Zeeman effect in helium-like ions, such as the second-order shift in magnetic field and the nuclear magnetic shielding. Typically, such problems are solved using perturbation theory. See, e.g., the PT calculations of the second-order Zeeman effect in hydrogen-like ions~\cite{manakov:74:jpb, manakov:76:pla, grozdanov:86:jpb, feinberg:90:pra, szmytkowski:02:jpb, szmytkowski:02:pra}, the second- and third-order Zeeman effects in boron-like ions~\cite{lindenfels:13:pra, glazov:13:ps, agababaev:17:nimb, varentsova:17:nimb, varentsova:18:pra}, and the Zeeman splitting in hydrogen-, lithium-, and boron-like ions with nonzero nuclear spin \cite{moskovkin:04:pra, moskovkin:06:pra, yerokhin:11:prl, yerokhin:12:pra, moskovkin:08:os, moskovkin:08:pra, volchkova:17:nimb}. Here we demonstrate application of the non-perturbative A-DKB method and compare the results with the PT calculations. This approach has been already applied by our group for the leading one-electron contributions~\cite{varentsova:17:nimb,volchkova:17:nimb}. We extend the calculations to the two-electron systems and compute the first-order electron-electron interaction, i.e., the one-photon-exchange contribution. The corresponding formulae for the matrix elements are given in the previous section. Below we concretize the potential $\hat{V}$ in Eq.~(\ref{eq:hD}) and give the formulae and numerical results for the contribution under consideration.
% present calculations for nonlinear contributions to the Zeeman effect for helium-like ions using the method. The application of the A-DKB method greatly simplifies the calculations.
% , i.e., the hyperfine-interaction correction to the $g$ factor
% 
\subsection {Quadratic contribution to the Zeeman effect}
\label{sec:g2}
Consider the bound electron in the presence of external magnetic field described by the stationary Dirac equation~(\ref{eq:D}) with $\hat{V}=\Vnucl(r)+\Vmagn$, where the spherically symmetric electrostatic nuclear potential $\Vnucl(r)$ is conveniently relegated to the zeroth-order Hamiltonian,
% 
% \begin{equation}
% \label{eq:DZ} 
%   \hat H \psi(\vec r) = E \psi(\vec r) \;,
% \end{equation}
% \begin{equation}
% \hat H=\hat H_0 + \hat H_{\mathrm{magn}}\;,
% \end{equation}
% where
\begin{equation}
\label{eq:H0}
  \hat{H_0} = \balpha \cdot \bfp + \beta + \Vnucl(r)
\;,
\end{equation}
while the magnetic-field interaction
\begin{equation}
\label{eq:Vmagn}
  \Vmagn = \lambda U
\;,\qquad
  \lambda = \muB B 
\;,\qquad
  U=[\bfr \times \balpha]_z
\;,
\end{equation}
is considered as a perturbation. Zeroth-order problem is spherically symmetric, the energies and the radial wave functions can be found numerically, e.g., within the DKB method~\cite{shabaev:04:prl}.
%  --- dimensionless parameter (in relativistic units).\\
Within the perturbation theory, the energy $E(\lambda)$ can be expanded in a power series in $\lambda$,
\begin{align}
\label{eq:E}
  E(\lambda) &= E^{(0)}+\Delta E^{(1)}+ \Delta E^{(2)}+...
\nonumber\\
    &= E^{(0)}+\lambda g^{(1)}+ \lambda^2 g^{(2)}+...
\;.
\end{align}
The first-order term, conveniently expressed via the $g$ factor,
\begin{equation}
  E^{(1)} = \lambda g^{(1)}(\MJ) = \muB B g \MJ
\;,
\end{equation}
and the second-order term, related to the magnetic susceptibility,
\begin{equation}
  E^{(2)} = \lambda^2 g^{(2)}(\MJ) = (\muB B)^2 g^{(2)}(\MJ)
\;,
\end{equation}
can be evaluated by the well-known perturbation theory formulae. We perform these calculations here for the $1s$ state using the previously developed numerical approach~\cite{agababaev:17:nimb,varentsova:18:pra}.

At the same time, the solutions of the Dirac equation can be found at a particular value of $\lambda$ within the A-DKB method described in Sec.~\ref{sec:A-DKB}.
% ?
Assuming that $E(\lambda)$ is known, the coefficients $g^{(1)}$ and $g^{(2)}$ can be found by differentiation with respect to $\lambda$ at the point $\lambda=0$:
\begin{align}
\label{eq:g1}
  g^{(1)} &= g \MJ = \left.{\frac{\displaystyle \partial E(\lambda)}{\displaystyle \partial{\lambda}}}\right|_{\lambda = 0} 
\;,\\
\label{eq:g2}
  g^{(2)} &= \left.{\frac{1}{2}\frac{\displaystyle {\partial}^2E(\lambda)}{\displaystyle \partial\lambda ^2}}\right|_{\lambda = 0}
\;.
\end{align}

Next, we consider the ground state of He-like ions and evaluate the electron-electron interaction contribution. The linear Zeeman effect is obviously zero in this case. The second-order coefficient is given by the second derivative of Eq.~(\ref{eq:Egs}),
\begin{align}
\label{eq:g2-gs}
  g^{(2)}[(1s)^2] = g^{(2)}_0[(1s)^2] + \Delta g^{(2)}_\mathrm{1ph}[(1s)^2]
\;,
\end{align}
where $g^{(2)}_0[(1s)^2]=2\,g^{(2)}[1s]$ is the value for non-interacting electrons, since $g^{(2)}(\MJ)=g^{(2)}(-\MJ)$.
To find the one-photon-exchange corrections to $g^{(2)}$, we can calculate the one-photon-exchange correction to the energy (\ref{eq:E_1ph}), and take the corresponding derivative,
\begin{align}
\label{eq:g2-1ph}
%   \Delta g^{(1)}_\mathrm{1ph} = \left.{\frac{d}{d\lambda}}\right|_{\lambda=0} \Delta E_\mathrm{1ph}(\lambda)
% \;,\\
  \Delta g^{(2)}_\mathrm{1ph} = \frac{1}{2} \left.{\frac{d^2}{d\lambda^2}}\right|_{\lambda=0} \Delta E_\mathrm{1ph}(\lambda)
\;.
\end{align}
% 
% For helium-like ion in the ground state the total $g$ factor is obviously zero.
% Then the total value of $g^{(2)}$ for helium-like ion is given by
% % 
% \begin{align}
%   g^{(2)} = 2\,g^{(2)} + \Delta g^{(2)}_\mathrm{1ph}
% \;,
% \end{align}
% % 

We evaluate the $g^{(1)}$ and $g^{(2)}$ for the $1s$ state and $\Delta g^{(2)}_\mathrm{1ph}$ for the $(1s)^2$ state according to the equations~(\ref{eq:g1}), (\ref{eq:g2}), and (\ref{eq:g2-1ph}). In order to find the derivatives, we solve the Dirac equation at the sets of $\lambda = -n\lambda_0, \dots, 0, \dots, n\lambda_0$ with $n=1,2,3$ and use the standard formulae for the derivatives. 
% 
% accomplish these calculation numerically
The key aspect of numerical differentiation is the choice of the optimal $\lambda_0$, which provides best accuracy. The error increases both for larger $\lambda_0$ (due to the approximate nature of the finite difference formulae) and for smaller $\lambda_0$ (due to machine accuracy limitations). Therefore, one has to make a series of calculations for a set of $\lambda_0$ and choose the optimal value in each case. For the calculations presented in this paper, we use $\lambda_0$ in the range $10^{-4}\dots 10^{-7}$.
As an independent test, we calculate $g^{(1)}[{1s}]$, $g^{(2)}[{1s}]$, and $\Delta g^{(2)}_\mathrm{1ph}[(1s)^2]$ within the standard perturbation theory~\cite{agababaev:17:nimb,varentsova:18:pra}. The results of these calculations are presented in Table~\ref{tab:1ph_g2}, very good agreement is observed in general.

\subsection{Zeeman splitting of the hyperfine structure levels}
\label{sec:sigma}

Rigorous consideration of the hyperfine interaction requires inclusion of the nuclear variables in the Hamiltonian. However, once the proper states of the system with a definite value of the total angular momentum $\bfF=\bfJ+\bfI$ are constructed, one can describe the hyperfine interaction by an effective term in the Dirac Hamiltonian for the bound electron,
\begin{equation}
\label{eq:Vhfs}
  \Vhfs = \mu W
\;,\qquad
%   \mu = \frac{1}{2\pi}\muB\mu_n
  \mu = \frac{\alpha}{2}\,\frac{m_e}{m_p}\,g_I I
\;,\qquad
  W = \frac{[\bfr\times\balpha]_z}{r^3}
\;,
\end{equation}
% 
%  $\mu_n = \muN g_I I$
where $I$ and $g_I$ are the nuclear spin and $g$ factor, respectively. We note that $\mu$ is the dimensionless parameter, proportional but not equal to the nuclear magnetic moment.
% it is not it. 
%  $\mu_n$
% where $\mu_n$ is the nuclear magnetic moment, see e.g.~\cite{moskovkin:04:pra,yerokhin:12:pra} for more details. 
% We use $\lambda$ and $\mu$ to ``tune'' the magnetic and hyperfine interactions, respectively.
Zeeman splitting of the hyperfine levels is then described by $\Vmagn$ (\ref{eq:Vmagn}). The total potential $\hat V$ in the Dirac equation is $\hat{V}=\Vnucl(r)+\Vmagn+\Vhfs$. From the perturbation-theory point of view, the zeroth-order Hamiltonian is given again by Eq.~(\ref{eq:H0}) and there are two perturbations, $\Vmagn$ and $\Vhfs$, characterized by two independent parameters, $\lambda$ and $\mu$, which ``tune'' the magnetic and hyperfine interactions, respectively.

In this work, we consider the weak-magnetic-field limit, when the Zeeman splitting is much smaller than the hyperfine splitting. In this case, the linear Zeeman shift is described by the $g$ factor of the electron-nucleus system, which can be written as~\cite{moskovkin:04:pra,yerokhin:12:pra},
\begin{align}
\label{eq:g_F}
%   g_F = g_J \frac{\langle \bfJ\cdot\bfF \rangle}{\langle F^2\rangle} - (1-\sigma)\frac{m_e}{m_p}g_I\frac{\langle \bfI\cdot\bfF \rangle}{\langle F^2\rangle}
  g_F &= g_J \frac{F(F+1)-I(I+1)+J(J+1)}{2F(F+1)} 
\nonumber\\
      &- (1-\sigma) \frac{m_e}{m_p} g_I \frac{F(F+1)+I(I+1)-J(J+1)}{2F(F+1)}
\;,
\end{align}
where $g_J$ is the electronic $g$ factor, $m_e$ and $m_p$ are the masses of electron and proton, respectively.
% \[
% \frac{\langle \vec I \vec F \rangle}{\langle F^2\rangle}=\frac{F(F+1)+I(I+1)-j(j+1)}{2F(F+1)}\;, \quad \frac{\langle \vec j \vec F \rangle}{\langle F^2\rangle} = \frac{F(F+1)+j(j+1)-I(I+1)}{2F(F+1)}\;.
% \]
% The formula (\ref{eq:g_F}) can be written as:
% \begin{equation}
% g_F=g_j\frac{\langle \vec j \vec F \rangle}{\langle F^2\rangle}  - (1-\sigma)\frac{m_e}{m_p}g_I\frac{\langle \vec I \vec F \rangle}{\langle F^2\rangle}\;.
% \end{equation}
The so-called nuclear magnetic shielding constant $\sigma$ corresponds to the mixed second-order contribution in $\Vmagn$ and $\Vhfs$.
% In the one-electron case of it is given by,
% % 
% \begin{equation}
% \label{eq:sigma}
%   \sigma = \alpha \sum_n \frac{\bra{a} W\ket{n}\bra{n} U\ket{a}}{\epsilon_a-\epsilon_n}
% \;,
% \end{equation}
% % 
% where $U$ and $W$ are given by Eqs.~(\ref{eq:Vmagn}) and (\ref{eq:Vhfs}), respectively. 
We calculate $\sigma$ for the $1s$ state within the perturbation theory following our previous work~\cite{volchkova:17:nimb}. Alternatively, this term can be found from the solutions of the Dirac equation~(\ref{eq:D}) with $\Vmagn$ and $\Vhfs$ included, by taking the mixed derivative,
\begin{equation}
\label{eq:sigma}
  \sigma =\frac{\alpha}{2}\left.{\frac{d^2}{d\lambda d\mu}}\right|_{\lambda=0, \mu =0} {E}{(\lambda, \mu)}
\;.
\end{equation}
For the sake of thorough and comprehensive check of the nonperturbative method, we also consider the hyperfine splitting itself within both A-DKB and PT methods.
% The hyperfine splitting 
In the one-electron approximation it is conveniently characterized by the relativistic factor $A(\alpha Z)$ (see, e.g., Refs.~\cite{shabaev:94:jpb,volotka:08:pra}) which can be found either as,
\begin{equation}
\label{eq:A-hfs}
  A = \frac{3}{8(\alpha Z)^3\MJ}\,\left.{\frac{\displaystyle \partial  E(\lambda, \mu)}{\displaystyle \partial{\mu}}}\right|_{\lambda=0, \mu =0}
\;,
\end{equation}
or within the first-order PT by replacing the derivative in Eq.~(\ref{eq:A-hfs}) with the average value of $W$.

For the ground state of He-like ions we write $\sigma[(1s)^2]$ as,
\begin{align}
\label{eq:sigma-gs}
  \sigma[(1s)^2] = \sigma_0[(1s)^2] + \Delta \sigma_\mathrm{1ph}[(1s)^2]
\;,
\end{align}
where $\sigma_0[(1s)^2]=2\,\sigma[1s]$. Note that the first term in Eq.~(\ref{eq:g_F}) vanishes since $J=0$ in this case.
The one-photon-exchange correction $\Delta \sigma_\mathrm{1ph}[(1s)^2]$ can be found from the corresponding energy correction~(\ref{eq:E_1ph}) by taking the derivatives,
\begin{equation}
\label{eq:sigma-1ph}
  \Delta \sigma_\mathrm{1ph} = \frac{\alpha}{2}\left.{\frac{d^2}{d\lambda d\mu}}\right|_{\lambda=0, \mu =0} \Delta E_\mathrm{1ph}(\lambda, \mu)
\;.
\end{equation}

We present the numerical results for the hyperfine splitting and the nuclear magnetic shielding constant in Table~\ref{tab:1ph_sigma}. The results obtained by differentiation of the energy with respect to $\lambda$ and $\mu$ (Eqs.~(\ref{eq:sigma}), (\ref{eq:A-hfs}), and (\ref{eq:sigma-1ph})) are labeled as A-DKB.  For the calculations presented in this paper, we use $\lambda_0$ in the range $10^{-4}\dots 10^{-6}$ and $\mu_0$ in the range $10^{-3}\dots 10^{-6}$. 
For the hyperfine-splitting factor $A$, the PT value obtained in this work and the one from Ref.~\cite{volotka:08:pra} are presented for comparison. We note that the finite nuclear size is taken into account in our calculations, while the finite nuclear magnetic moment distribution (the Bohr-Weisskopf effect) is not. So, we actually give the value of $A(\alpha Z)(1-\delta)$ here, where $\delta$ is the nuclear size correction~\cite{shabaev:94:jpb,volotka:08:pra}.
Both $\sigma[1s]$ and $\Delta \sigma_\mathrm{1ph}[(1s)^2]$ are also evaluated within the standard perturbation theory following the same procedure as for $g^{(2)}$. These values are labeled as PT in Table~\ref{tab:1ph_sigma}.
% For comparison, we present the results obtained within the standard PT approach. 
The values of $\sigma[1s]$ from Ref.~\cite{moskovkin:04:pra} are presented for comparison as well. The agreement between these independent results serves as a robust test of the proposed approach based on the A-DKB solutions of the Dirac equation in the presence of external magnetic field and hyperfine interaction.

\subsection{Discussion}

The results presented in Tables~\ref{tab:1ph_g2} and \ref{tab:1ph_sigma} demonstrate the agreement between the nonperturbative approach and the standard PT. The small deviations observed are due to the limited basis set of the A-DKB method, which can be increased when needed. We note that within the PT framework the one-photon-exchange contributions, $\Delta g^{(2)}_\mathrm{1ph}[(1s)^2]$ and $\Delta \sigma_\mathrm{1ph}[(1s)^2]$, comprise several types of diagrams and rather bulky set of formulae, see, e.g., the detailed derivation for the nuclear magnetic shielding in Refs.~\cite{moskovkin:08:os,moskovkin:08:pra}. Within the A-DKB approach we only need to calculate $\DEph$ (\ref{eq:E_1ph}) at the appropriate set of $\lambda$ and $\mu$ values. At this level, the A-DKB and PT methods are comparable in efficiency. However, for any higher-order contributions (e.g., third order in magnetic field, second order in the interelectronic interaction, or especially QED corrections) the use of PT will be almost prohibitively expensive in terms of complexity and computational burden. In Ref.~\cite{varentsova:18:pra} we have tackled the one-photon-exchange corrections to the second- and third-order Zeeman effect, $\Delta g^{(2)}_\mathrm{1ph}$ and $\Delta g^{(3)}_\mathrm{1ph}$. The method that we used there is based on the recursive perturbation theory (RPT) with respect to $\Vmagn$. RPT was used initially to reach higher-order terms of Zeeman and Stark shifts in Ref.~\cite{rozenbaum:14:pra} and generalized to the electron-electron interaction operator within the Breit approximation in Ref.~\cite{glazov:17:nimb}. Still, the method of Ref.~\cite{varentsova:18:pra} requires ``accumulation'' of the PT contributions at finite $\lambda$ and numerical differentiation with respect to $\lambda$. The results obtained there for $\Delta g^{(3)}_\mathrm{1ph}$ were not reproduced within the standard PT due to the combinatorial and computational complexity mentioned above. To summarize, we consider the proposed approach as the most suitable one for the higher-order calculations needed to achieve accurate theoretical predictions.
% The A-DKB method would require only 

% \section{Results}
% All integrals were taken by the Gauss method. Differentiation was carried out using the finite difference method. The main difficulty of numerical differentiation is the selection of the optimal step of differentiation $h$, providing maximum accuracy. The differentiation error can increase both with increasing step $h$ (caused by inaccuracies in the approximation of the function), and with its decrease (caused by machine accuracy limitations). Therefore, the selection of the optimal step $h$ is an important task. Here we use $\lambda$ in the range ... .\\
% The results for the quadratic Zeeman effect are shown in table \ref{tab:1ph_g2}, for the nuclear magnetic shielding constant in table \ref{tab:1ph_sigma}. To verify A-DKB method, we compare the A-DKB method with the perturbation theory.

\section{Conclusion}

The nonperturbative A-DKB method to solve the Dirac equation in the axially symmetric field has been applied to evaluation of the Zeeman effect in helium-like ions. The second-order contribution in magnetic field and the nuclear magnetic shielding constant are calculated for the ground $(1s)^2$ state by differentiation of the energy with respect to the magnetic-field and hyperfine interaction parameters. The first-order contribution of the electron-electron interaction is taken into account as well. All of these values are calculated within the standard perturbation theory, starting from the Dirac equation in the Coulomb field of the nucleus. The results of these two independent methods are found to be in good agreement. The presented method provides an efficient alternative to the perturbation theory, it is foreseen to be advantageous for higher-order calculations. The results for the nuclear magnetic shielding constant can be used for determination of the nuclear magnetic moments. The quadratic Zeeman effect can be relevant for the high-precision Penning-trap measurements of the transition energies in He-like ions. To provide the complete theoretical background, the corresponding calculations for excited states are in demand which is the subject of our future investigations. 

% \section{Acknowledgements}
\acknowledgments
We thank Artem Kotov for valuable discussions.
The work was supported by RFBR (Grants No. 19-32-90278 and 19-02-00974), by DFG (Grant No.~VO 1707/1-3), by the Foundation for the Advancement of Theoretical Physics and Mathematics ``BASIS'', and by the German–Russian Interdisciplinary Science Center (G-RISC).

\begin{table}
\caption{ \label{tab:1ph_g2} The $1s$-state $g$-factor and the second-order Zeeman effect for the ground $(1s)^2$ state of helium-like ions in terms of the $g^{(2)}$ coefficient --- the leading-order term $g^{(2)}_0$, the one-photon-exchange correction $\Delta g^{(2)}_\mathrm{1ph}$, and their sum $g^{(2)}$. A-DKB and PT denote the method of calculation, see text for details. The $g^{(2)}$ values are given in units of $10^{-3}$.}
\begin{center}
\begin{tabular}{rlllll}
% \hline
\hline
 $Z$
& method 
& $g[1s]$
& $g^{(2)}_0[(1s)^2] \times 10^3$
& $\Delta g^{(2)}_\mathrm{1ph}[(1s)^2] \times 10^3$
& $g^{(2)}[(1s)^2] \times 10^3$
% & \multicolumn{3}{c} {$g^{(2)}$}\\
% \hline
% &
% & $g_0$
% & $g^{(2)}_0 \times 10^3$
% & $\Delta g^{(2)}_{1ph}\times 10^3$
% & $g^{(2)}_{1ph}\times 10^3$
\\
\hline
$6$  & A-DKB & $1.998721$ & $1.040617$ & $0.277574$ & $1.318190$ \\
     & PT    & $1.998721$ & $1.040605$ & $0.277487$ & $1.318092$\\
     \hline
$10$ & A-DKB & $1.996445$ & $0.372907$ & $0.060094$ & $0.433001$\\
     & PT    & $1.996445$ & $0.372913$ & $0.060075$ & $0.432988$ \\ \hline
$12$ & A-DKB & $1.994878$ & $0.258153$ & $0.034863$ & $0.293015$ \\
     & PT    & $1.994878$ & $0.258154$ & $0.034821$ & $0.292975$ \\ \hline  
$14$ & A-DKB & $1.993024$ & $0.188958$ & $0.021974$ & $0.210932$ \\
     & PT    & $1.993024$ & $0.188959$ & $0.021968$ & $0.210927$ \\ \hline
$16$ & A-DKB & $1.990881$ & $0.144048$ & $0.014752$ & $0.158800$ \\
     & PT    & $1.990881$ & $0.144049$ & $0.014748$ & $0.158797$ \\ \hline
$18$ & A-DKB & $1.988448$ & $0.113259$ & $0.010384$ & $0.123643$\\
     & PT    & $1.988448$ & $0.113259$ & $0.010383$ & $0.123642$\\ \hline
$20$ & A-DKB & $1.985723$ & $0.091236$ & $0.007589$ & $0.098825$ \\
     & PT    & $1.985723$ & $0.091236$ & $0.007589$ & $0.098825$ \\ \hline 
$32$ & A-DKB & $1.963138$ & $0.034033$ & $0.001889$ & $0.035921$\\
     & PT    & $1.963138$ & $0.034032$ & $0.001891$ & $0.035923$\\ \hline
\end{tabular}
\end{center}
\end{table}

\begin{table}

\caption{ \label{tab:1ph_sigma} The factor $A(\alpha Z)$ for state 1s and the nuclear magnetic shielding constant $\sigma$ for the ground $(1s)^2$ state of helium-like ions. The leading-order term $\sigma_0[(1s)^2]$, the one-photon-exchange correction $\Delta \sigma_\mathrm{1ph}[(1s)^2]$, and their sum $\sigma[(1s)^2]$. The $1s$ results from Ref.~\cite{moskovkin:04:pra} are given for comparison ($\sigma_0[(1s)^2]=2\sigma[1s]$). A-DKB and PT denote the method of calculation, see text for details. The values of $\sigma$ are given in units of $10^{-3}$.}
\begin{center}
\begin{tabular}{rlllll}
% \hline
\hline
 $Z$
& method
& $A(\alpha Z)$
& $\sigma_0[(1s)^2] \times 10^3$
& $\Delta \sigma_\mathrm{1ph}[(1s)^2]\times 10^3$
& $\sigma[(1s)^2]\times 10^3$
\\
\hline
$6$  & A-DKB & $1.002332$   & $0.214116$   & $-0.022093$ & $0.192022$ \\
     & PT    & $1.002332$   & $0.214108$   & $-0.022082$ & $0.192025$ \\ 
     &       &              & $0.214109^b$ &             &            \\\hline
$10$ & A-DKB & $1.006906$   & $0.360126$   & $-0.021893$ & $0.338233$ \\
     & PT    & $1.006906$   & $0.360134$   & $-0.021888$ & $0.338245$ \\
     &       & $1.006911^a$ & $0.360135^b$ &             &            \\\hline
$12$ & A-DKB & $1.010201$   & $0.434905$   & $-0.021767$ & $0.413138$ \\
     & PT    & $1.010201$   & $0.434893$   & $-0.021751$ & $0.413142$  \\
     &       & $1.010204^a$ &              &             &\\ \hline
$14$ & A-DKB & $1.014136$   & $0.511180$   & $-0.021601$ & $0.489578$ \\
     & PT    & $1.014136$   & $0.511171$   & $-0.021584$ & $0.489586$ \\
     &       & $1.014133^a$ &              &             &\\\hline
$16$ & A-DKB & $1.018689$   & $0.589227$   & $-0.021414$ & $0.567813$ \\
     & PT    & $1.018689$   & $0.589235$   & $-0.021386$ & $0.567849$ \\ 
     &       & $1.018686^a$ & $0.589241^b$ &             &            \\\hline
$18$ & A-DKB & $1.023920$   & $0.669409$   & $-0.021310$ & $0.648098$ \\
     & PT    & $1.023920$   & $0.669384$   & $-0.021154$ & $0.648231$ \\ \hline
$20$ & A-DKB & $1.029879$   & $0.751866$   & $-0.020821$ & $0.731045$ \\
     & PT    & $1.029879$   & $0.751898$   & $-0.020884$ & $0.731014$ \\
     &       & $1.029872^a$ & $0.751916^b$ &             &            \\\hline
$32$ & A-DKB & $1.081318$   & $1.315684$   & $-0.018614$ & $1.297070$ \\
     & PT    & $1.081318$   & $1.315699$   & $-0.018264$ & $1.297435$ \\
     &       &              & $1.315862^b$ &             &            \\\hline
\end{tabular}
\end{center}
$^a$From Ref. \cite{volotka:08:pra}, \qquad
$^b$From Ref. \cite{moskovkin:04:pra}
\end{table}


\begin{thebibliography}{99}
% 
\bibitem{sturm:17:a}
S.~Sturm, M.~Vogel, F.~K\"ohler-Langes, W.~Quint, K.~Blaum, and G.~Werth,
Atoms~\textbf{5}, 4~(2017).
% 
\bibitem{sturm:14:n}
% High-precision measurement of the atomic mass of the electron
S.~Sturm, F.~K\"ohler, J.~Zatorski, A.~Wagner, Z.~Harman, G.~Werth, W.~Quint, C.~H.~Keitel, and K.~Blaum,
Nature \textbf{506}, 467 (2014).
% 
\bibitem{glazov:19:prl}
% $g$ Factor of Lithiumlike Silicon: New Challenge to Bound-State QED
D. A. Glazov, F. K\"ohler-Langes, A. V. Volotka, K. Blaum, F. Hei\ss{}e, G. Plunien, W. Quint, S. Rau, V. M. Shabaev, S. Sturm, and G. Werth,
Phys. Rev. Lett. \textbf{123}, 173001 (2019).
%
\bibitem{arapoglou:19:prl}
% The g-factor of Boronlike Argon 40_Ar^13+
I.~Arapoglou, A.~Egl, M.~H\"ocker, T.~Sailer, B.~Tu, A.~Weigel, R.~Wolf, H.~Cakir, V.~A.~Yerokhin, N.~S.~Oreshkina, V.~A.~Agababaev, A.~V.~Volotka, D.~V.~Zinenko, D.~A.~Glazov, Z.~Harman, C.~H.~Keitel, S.~Sturm, and K.~Blaum,
Phys. Rev. Lett. \textbf{122}, 253001 (2019).
% %
% \bibitem{sturm:13:ap}
% % Electron g-factor determinations in Penning traps
% S.~Sturm, G.~Werth, and K.~Blaum,
% Ann. Phys. (Berlin) \textbf{525}, 620 (2013).
% %
% \bibitem{volotka:13:ap}
% % Progress in quantum electrodynamics theory of highly charged ions
% A.~V.~Volotka, D.~A.~Glazov, G.~Plunien, and V.~M.~Shabaev,
% Ann. Phys. (Berlin) \textbf{525}, 636 (2013).
%
\bibitem{shabaev:15:jpcrd}
% Theory of Bound-Electron g Factor in Highly Charged Ions
V.~M.~Shabaev, D.~A.~Glazov, G.~Plunien, and A.~V.~Volotka,
J.~Phys.~Chem.~Ref.~Data~\textbf{44}, 031205~(2015).
%
\bibitem{harman:18:jpcs}
Z.~Harman, B.~Sikora, V.~A.~Yerokhin, H.~Cakir, V.~Debierre, N.~Michel, N.~S.~Oreshkina, N.~A.~Belov, J.~Zatorski, and C.~H.~Keitel, 
J. Phys.: Conf. Ser. \textbf{1138}, 012002~(2018).
% 
% %11
% \bibitem{shabaev:06:prl}
% % g factor of heavy ions: a new access to the fine structure constant
% % V.~M.~Shabaev \textit{et al.},
% V.~M.~Shabaev, D.~A.~Glazov, N.~S.~Oreshkina, A.~V.~Volotka, G.~Plunien, H.-J.~Kluge, and W.~Quint,
% Phys.~Rev.~Lett.~\textbf{96}, 253002~(2006).
%12
\bibitem{werth:01:ha}
G.~Werth, H.~Häffner, N.~Hermanspahn, \mbox{H.-J.}~Kluge, W.~Quint, J.~Verd\'u, in:~S.~G.~Karshenboim~et~al.~(Eds.), The~Hydrogen~Atom, Springer, Berlin, 2001, p. 204.
%15
\bibitem{quint:08:pra}
W.~Quint, D.~Moskovkhin, V.~M.~Shabaev, and M.~Vogel,
Phys.~Rev.~A~\textbf{78}, 032517~(2008).
%13
\bibitem{ullmann:17:natcommun}
J.~Ullmann, Z.~Andelkovic, C.~Brandau \textit{et al.}, Nat.~Commun.~\textbf{8}, 15484~(2017).
%
\bibitem{volotka:12:prl}
% Test of Many-Electron QED Effects in the Hyperfine Splitting of Heavy High-Z Ions
A.~V.~Volotka, D.~A.~Glazov, O.~V.~Andreev, V.~M.~Shabaev, I.~I.~Tupitsyn, and G.~Plunien
Phys. Rev. Lett. \textbf{108}, 073001 (2012).
%14
\bibitem{skripnikov:18:prl}
L.~V.~Skripnikov, S.~Schmidt, J.~Ullmann, C.~Geppert, F.~Kraus, B.~Kresse, W.~N\"ortersh\"auser, A.~F.~Privalov, B.~Scheibe, V.~M.~Shabaev, M.~Vogel, and A.~V.~Volotka, 
Phys.~Rev.~Lett.~\textbf{120}, 093001~(2018).
%14
\bibitem{fella:20:prr}
% Magnetic moment of 207Pb and the hyperfine splitting of 207Pb81+
V. Fella, L. V. Skripnikov, W. N\"ortersh\"auser, M. R. Buchner, H. L. Deubner, F. Kraus, A. F. Privalov, V. M. Shabaev, and M. Vogel,
Phys. Rev. Research \textbf{2}, 013368 (2020).
%
\bibitem{moore:99:mp}
E.~A.~Moore,
Mol.~Phys.~\textbf{97}, 375~(1999).
%
\bibitem{pyper:99:mp1}
N.~C.~Pyper,
Mol.~Phys.~\textbf{97}, 381~(1999).
%
\bibitem{pyper:99:mp2}
N.~C.~Pyper and Z.~C.~Zhang,
Mol.~Phys.~\textbf{97}, 391~(1999).
%2
%5
\bibitem{moskovkin:04:pra}
D.~L.~Moskovkin, N.~S.~Oreshkina, V.~M.~Shabaev, T.~Beier, G.~Plunien, W.~Quint, and G.~Soff, 
Phys.~Rev.~A~\textbf{70}, 032105~(2004).
%6
\bibitem{moskovkin:06:pra}
D.~L.~Moskovkin and V.~M.~Shabaev,
Phys.~Rev.~A~\textbf{73}, 052506~(2006).
%7
\bibitem{yerokhin:11:prl}
V.~A.~Yerokhin, K.~Pachucki, Z.~Harman, and C.~H.~Keitel, Phys.~Rev.~Lett.~\textbf{107} 043004 (2011).
%8
\bibitem{yerokhin:12:pra}
V.~A.~Yerokhin, K.~Pachucki, Z.~Harman, and C.~H.~Keitel, Phys. Rev. A \textbf{85}, 022512 (2012).
%9
\bibitem{moskovkin:08:os}
D.~L.~Moskovkin, V.~M.~Shabaev, and W.~Quint, Opt.~Spectrosc.~\textbf{104}, 637~(2008).
%10
\bibitem{moskovkin:08:pra}
D.~L.~Moskovkin, V.~M.~Shabaev, and W.~Quint, 
Phys.~Rev.~A~\textbf{77}, 063421~(2008).
%
\bibitem{volchkova:17:nimb}
% Nuclear magnetic shielding in boronlike ions
A.~M.~Volchkova, A.~S.~Varentsova, N.~A.~Zubova, V.~A.~Agababaev, D.~A.~Glazov, A.~V.~Volotka, V.~M.~Shabaev, and G.~Plunien,
Nucl. Instrum. Methods Phys.~Res.~B~\textbf{408}, 89~(2017).
%
\bibitem{lindenfels:13:pra}
% Experimental access to higher-order Zeeman effects by precision spectroscopy of highly charged ions in a Penning trap
D.~von~Lindenfels, M.~Wiesel, D.~A.~Glazov, A.~V.~Volotka, M.~M.~Sokolov, V.~M.~Shabaev, G.~Plunien, W.~Quint, G.~Birkl, A.~Martin, and M.~Vogel,
Phys. Rev. A \textbf{87}, 023412 (2013).
%
\bibitem{glazov:13:ps}
% g factor of boron-like ions: ground and excited states
D.~A.~Glazov, A.~V.~Volotka, A.~A.~Schepetnov, M.~M.~Sokolov, V.~M.~Shabaev, I.~I.~Tupitsyn, and G.~Plunien,
Phys. Scr. \textbf{T156}, 014014 (2013).
%4
\bibitem{varentsova:17:nimb}
A.~S.~Varentsova, V.~A.~Agababaev, A.~M.~Volchkova, D.~A.~Glazov, A.~V.~Volotka, V.~M.~Shabaev, and G.~Plunien,
Nucl. Instum. Methods Phys.~Res.~B~\textbf{408}, 80~(2017).
%4
\bibitem{varentsova:18:pra}
A.~S.~Varentsova, V.~A.~Agababaev, D.~A.~Glazov, A.~M.~Volchkova, A.~V.~Volotka, V.~M.~Shabaev, and G.~Plunien, 
Phys.~Rev.~A~\textbf{97}, 043402~(2018).
%3
\bibitem{agababaev:17:nimb}
V.~A.~Agababaev, A.~M.~Volchkova, A.~S.~Varentsova, D.~A.~Glazov, A.~V.~Volotka, V.~M.~Shabaev and G.~Plunien,
Nucl.~Instrum.~Methods~Phys.~Res.~B~\textbf{408}, 70~(2017).
%3
\bibitem{agababaev:20}
V.~A.~Agababaev \etal, 
% A.~M.~Volchkova, A.~S.~Varentsova, D.~A.~Glazov, A.~V.~Volotka, V.~M.~Shabaev and G.~Plunien,
to be published.
% 
\bibitem{egl:19:prl}
% Application of the Continuous Stern-Gerlach Effect for Laser Spectroscopy of the $^{40}{\mathrm{Ar}}^{13+}$ Fine Structure in a Penning Trap
A. Egl, I. Arapoglou, M. H\"ocker, K. K\"onig, T. Ratajczyk, T. Sailer, B. Tu, A. Weigel, K. Blaum, W. N\"ortersh\"auser, and S. Sturm,
Phys. Rev. Lett. \textbf{123}, 123001 (2019).
% 
\bibitem{epp:15:pra}
S. W. Epp, R. Steinbr\"ugge, S. Bernitt, J. K. Rudolph, C. Beilmann, H. Bekker, A. M\"uller, O. O. Versolato, H.-C. Wille, H. Yavas, J. Ullrich, and J. R. Crespo L\'opez-Urrutia, 
Phys. Rev. A \textbf{92}, 020502(R) (2015).
% 
\bibitem{beiersdorfer:15:pra}
P. Beiersdorfer and G. V. Brown, 
Phys. Rev. A \textbf{91}, 032514 (2015).
% 
\bibitem{machado:18:pra}
J. Machado, C. I. Szabo, J. P. Santos, P. Amaro, M. Guerra, A. Gumberidze, G. Bian, J. M. Isac, and P. Indelicato, 
Phys. Rev. A \textbf{97}, 032517 (2018).
% 
\bibitem{malyshev:19:pra}
A.~V.~Malyshev, Y.~S.~Kozhedub, D.~A.~Glazov, I.~I.~Tupitsyn, and V.~M.~Shabaev,
Phys. Rev. A \textbf{99}, 010501(R) (2019).
% 
\bibitem{kozhedub:19:pra}
Y.~S.~Kozhedub, A.~V.~Malyshev, D.~A.~Glazov, V.~M.~Shabaev, and I.~I.~Tupitsyn,
Phys. Rev. A \textbf{100}, 062506 (2019).
% 
\bibitem{yerokhin:19:jpcrd}
% Theoretical Energy Levels of 1 sns and 1 snp States of Helium-Like Ions
V.~A.~Yerokhin and A.~Surzhykov,
J.~Phys.~Chem.~Ref.~Data~\textbf{48}, 033104 (2019).
% 
\bibitem{rozenbaum:14:pra}
E.~B.~Rozenbaum, D.~A.~Glazov, V.~M.~Shabaev, K.~E.~Sosnova, and D.~A.~Telnov, 
Phys.~Rev.~A~\textbf{89}, 012514~(2014).
%1
\bibitem{shabaev:04:prl}
% Dual Kinetic Balance Approach to Basis-Set Expansions for the Dirac Equation
V.~M.~Shabaev, I.~I.~Tupitsyn, V.~A.~Yerokhin, G.~Plunien, and G.~Soff,
Phys.~Rev.~Lett.~\textbf{93}, 130405~(2004).
%
\bibitem{akhiezer}
A. I. Akhiezer and V. B. Berestetsky,
\textit{Quantum Electrodynamics}, New York (1965).
% 
\bibitem{chen:92:pra}
% Relativistic and nonrelativistic finite-basis-set calculations of low-lying levels of hydrogenic atoms in intense magnetic fields
Z.~Chen and S.~P.~Goldman, 
Phys.~Rev.~A \textbf{45}, 1722 (1992).
%
\bibitem{rutkowski:05:ps}
% Analytical Solution for Relativistic Hydrogenic Atom in Static and Uniform Magnetic Field
A.~Rutkowski and A.~Poszwa,
Phys.~Scr.~\textbf{71}, 484~(2005).
%
\bibitem{nakashima:10:aj}
% Solving the Schr\"odinger and Dirac Equations for a Hydrogen Atom in the Universe's Strongest Magnetic Fields With the Free Complement Method
H.~Nakashima and H.~Nakatsuji,
Astrophys.~J.~\textbf{725}, 528~(2010).
%
\bibitem{manakov:74:jpb}
% Relativistic electromagnetic susceptibilities of hydrogen-like atoms
N.~L.~Manakov, L.~P.~Rapoport, and S.~A.~Zapryagaev,
J.~Phys.~B \textbf{7}, 1076 (1974).
%
\bibitem{manakov:76:pla}
% A reduced green function of the Dirac equation with a coulomb potential. Second order Zeeman effect
N.~L.~Manakov and S.~A.~Zapryagaev,
Phys.~Lett.~A~\textbf{58}, 23~(1976).
%
\bibitem{grozdanov:86:jpb}
% Second-order perturbation calculations for the hydrogenic Zeeman effect
T.~P.~Grozdanov and H.~S.~Taylor,
J. Phys. B \textbf{19}, 4075 (1986).
%
\bibitem{feinberg:90:pra} 
% Quadratic Zeeman effect in positronium
G.~Feinberg, A.~Rich, and J.~Sucher, 
Phys.~Rev.~A \textbf{41}, 3478~(1990).
%
\bibitem{szmytkowski:02:jpb}
% Magnetizability of the relativistic hydrogen-like atom: application of the Sturmian expansion of the first-order Dirac–Coulomb Green function
R.~Szmytkowski,
J. Phys. B \textbf{35}, 1379 (2002).
%
\bibitem{szmytkowski:02:pra}
% Larmor diamagnetism and Van Vleck paramagnetism in relativistic quantum theory: The Gordon decomposition approach
R.~Szmytkowski,
Phys.~Rev.~A~\textbf{65}, 032112 (2002).
% ; J. Phys. B \textbf{35}, 1379 (2002).
%
\bibitem{shabaev:94:jpb}
% Hyperfine structure of hydrogen-like ions
V.~M.~Shabaev,
J. Phys. B \textbf{27}, 5825 (1994).
% 
\bibitem{volotka:08:pra}
%Ground-state hyperfine structure of H-, Li-, and B-like ions in the intermediate-Z region
A.~V.~Volotka, D.~A.~Glazov, I.~I.~Tupitsyn, N.~S.~Oreshkina, G.~Plunien, and V.~M.~Shabaev,
Phys.~Rev.~A~\textbf{78}, 062507 (2008).
% 
\bibitem{glazov:17:nimb}
% Higher-order perturbative relativistic calculations for few-electron atoms and ions
D.~A.~Glazov, A.~V.~Malyshev, A.~V.~Volotka, V.~M.~Shabaev, I.~I.~Tupitsyn, and G.~Plunien,
Nucl. Instr. Meth. Phys. Res.~B \textbf{408}, 46~(2017).
% 
\end{thebibliography}
\end{document}